# Density Fluctuation in the Tandem Mirror GAMMA 10

A. Itakura, S. Tsunoda, M. Fukuhara, H. Higaki, H. Hojo, M. Ichimura, K. Ishii,

Y. Shima, H. Takiue, M. Yoshikawa, T. Cho

*Plasma Research Center, University of Tsukuba, Tsukuba, 305-8577, Japan*

The tandem mirror GAMMA 10 utilizes an electron cyclotron resonance heating (ECRH) for forming a confinement potential. Density fluctuation is observed using microwaves, such as interferometer, reflectometry and Fraunhofer diffraction (FD) method. An ultrashort-pulse reflectometry has an advantage of detecting fluctuation locally. The wave number can be obtained by the FD method. In the edge plasma region electrostatic probes are used. Fluctuation with coherent mode in several kHz is excited in the hot-ion mode plasma. When the ECRH is applied, electron density in the central cell increases gradually. On the contrary, intensity of the density fluctuation decreases and coherent mode is suppressed. Its frequency also varies with the density increase. In that time, radial potential distribution, i.e., electric field, changes along with the formation of plug potential. From this behaviour it is deduced that the fluctuation deeply relates to the potential formation and improvement of the confinement.

1. Introduction

The GAMMA 10 is the axisymmetrized tandem mirror with a thermal barrier [1]. Plasma is confined by an electric plug potential along with a magnetic mirror. Plug potential is created by an electron cyclotron resonance heating (ECRH) and thermal barrier potential is also created by the second harmonics of the ECRH. In this experiment, GAMMA 10 is operated in the hot ion mode using an ion cyclotron range of frequency (ICRF) [2]. The recent status of the GAMMA 10 is described in ref. 3.

Fluctuation of several kHz in the electron density is observed in the central cell. When the plug ECRH is applied, central cell density increases with increase of plug potential. On the contrary, fluctuation of density in coherent mode is suppressed in that time. Frequency also changes its value. Observation by microwave and by electrostatic probes in the edge region show somewhat different results. So it suggests that some nonlinear effects happen in the plasma.

2. Experimental Arrangement

Detailed description of experimental apparatus of the GAMMA 10 is given in ref.3. Electron density is measured using microwave interferometers, whose frequency is 70 GHz. Density fluctuation is observed by applying Fast Fourier Transformation to the output of the interferometers. A Fraunhofer-diffraction (FD) method is installed to observe fluctuation with wave number [4]. An ultra-short pulse reflectometry is also utilized to observe information of density [5]. Time of flight of the reflected wave has local information of density fluctuation at the cutoff layer. Frequency of incident wave is ranged from 7 GHz to 11 GHz. It corresponds to the density of 0.6 to 1.5 x $10^{18}$ m$^{-3}$. In the edge region of the central cell, electrostatic probes are installed [6]. Density and floating potential and their fluctuations are observed.

3. Experimental Results and Discussions

In the hot ion mode, plasma is produced at 50.5 ms and sustatined by ICRF. Then ECRH is applied to create confining potential and thermal barrier potential. The density in the central cell is measured by the interferometer. A time variation of line-integrated density is shown in Fig.1 along with the variation of energy density and time sequence of heating system.

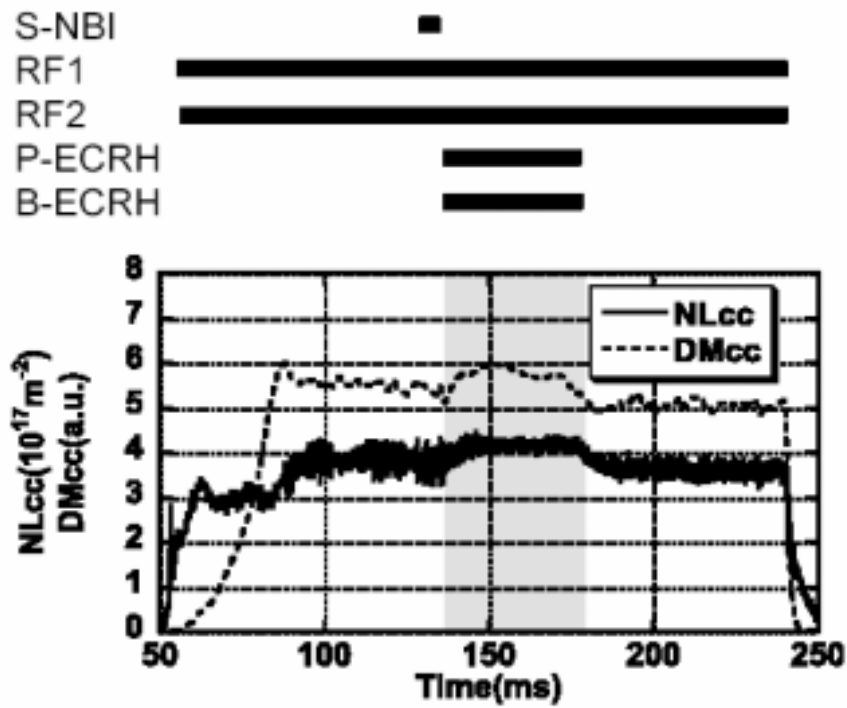

Fig.1 Time variation of line-integrated density (NLcc) and
energy density (DMcc) with injection time of heating systems.

Fluctuation measured by FD method is shown in Fig.2 (a). Two modes of signal is expressed. One is initially about 9 kHz. When ECRH is applied, frequency decreases and intensity decreases The second one is initially about 3 kHz. When ECRH is applied, intensity also decreases but frequency increses. Figure 2 (b) showds fluctuations measured by the electrostatic probein the edge region. This expresses also two mode. Higher frequency mode shows same behaviour of frequency variation and intensity as the measured by FD method.

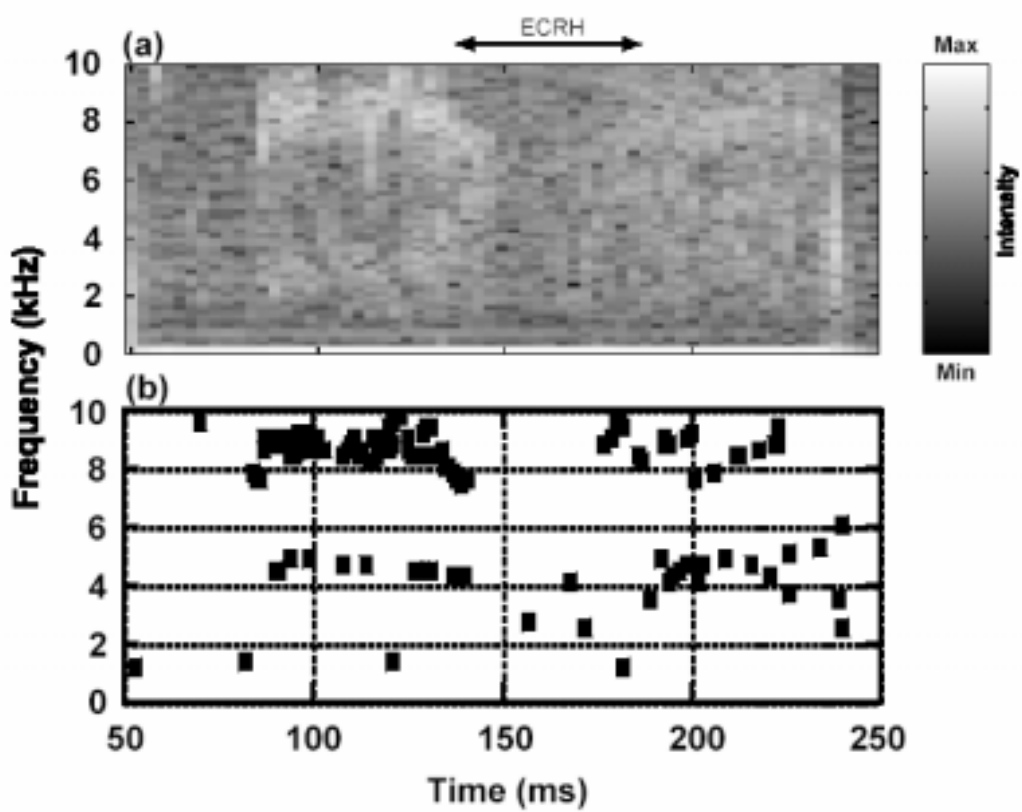

Fig.2 Fluctuations (a) measured by Fraunhofer-diffraction method
and (b) measured by the electrostatic probe.

But, Lower mode doesnot. It frequency is about a half of the higher one. Although the FD method diagnoses through the core plasma, the electrostatic probe measures only in the edge region. This phenomennon suggests that in the core region or near the core region there are two modes of fluctuations and some nonlinear mechanisu occours.

In the central cell, heavy ion beam probe for potential measurement is installed. Figure3 shows the potential on the axis $\Phi_c$ and also floating potential at the edge measure by the limiter. This figure shows that the potential on the axis increases with application of ECRH, i.e., formation of plug potential. So direction of the electric field varies invertly. This causes the direction of E x B drift. Then it causes the frequency change and suppression of fluctuations.

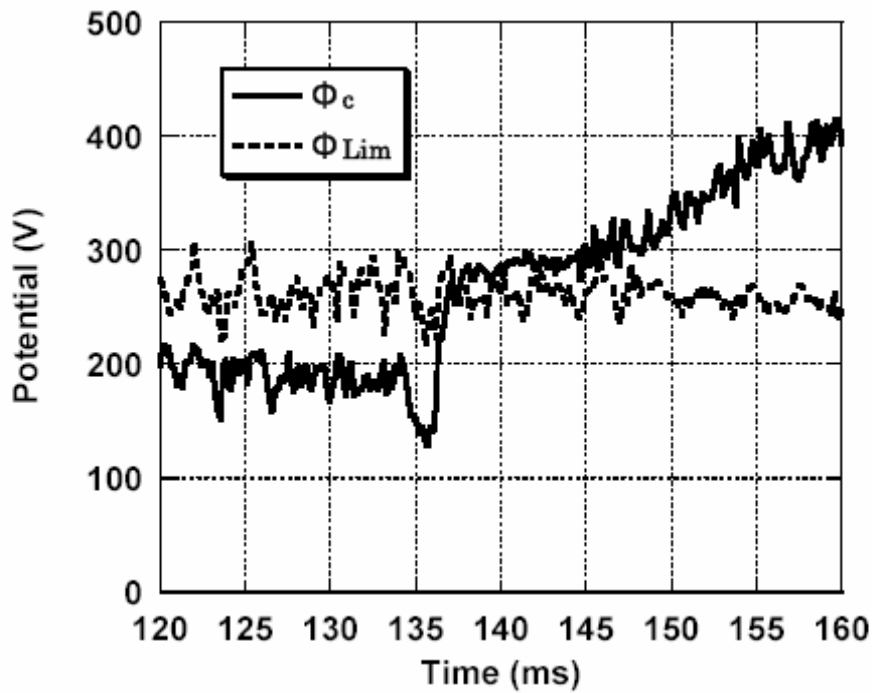

Fig.3 Potential on the axis measured by the heavy ion beam probe $\Phi_c$ and floating potential at the edge measured by limiter $\Phi_{Lim}$.

4. Summary

When the ECRH is applied for creation of confining potential, direction of electric field in the central cell plasma inverts. Intensity of fluctuation decreses. But fluctuations in the inner part of plasma and in the edge region plasma show different behaviour. It suggests that there is some nonlinear phenomenon in the inner part of plasma.

Acknowledgements

The authors deeply acknowledge the collaboration of the member of GAMMA 10 group, University of tsukuba.